\documentclass[aps,prl,twocolumn,floatfix]{revtex4}
\usepackage{graphicx}

\begin{document}

\title{Reaching $^7$Li BEC with a Mini-Trap}

\author{Ruquan Wang}
\author{Mingchang Liu}
\author{Francesco Minardi}
\altaffiliation{CNR-INFM, LENS, Via Nello Carrara, 1 50019 Sesto
Fiorentino}
\author{Mark Kasevich}
\email{kasevich@stanford.edu}
\affiliation{Department of Physics, Stanford University, Stanford CA, 94305}%

\date{\today}

\begin{abstract}
A novel mm-scale Ioffe-Pritchard trap is used to achieve
Bose-Einstein condensation in $^7$Li.  The trap employs
free-standing copper coils integrated onto a direct-bond copper
surface electrode structure. The trap achieves a radial magnetic
gradient of 420 G/cm, an axial oscillation frequency of 50 Hz and
a trap depth of 66 G with a 100 A drive current and 7 W total
power dissipation.
\end{abstract}

\pacs{39.90.+d, 39.20.+q, 39.10.+j} \maketitle

\section{Introduction}
Forced evaporative cooling in conservative magnetic traps has
become a key technology to cool atoms to quantum degeneracy.  In
1995, the invention of the time-orbiting potential (TOP) trap
paved the way to the first Bose-Einstein condensate
(BEC)~\cite{Rb-BEC}. In the following years, the cloverleaf
trap~\cite{clover}, the QUIC trap~\cite{quic}, and other
conservative traps allowed further improvements in trap
performance.  More recently, magnetic traps have been realized
with micro-fabricated wires on a surface (atom
chips)~\cite{atomchip1,atomchip2}.

In this paper we report on the realization of a novel
Ioffe-Pritchard \cite{Pethick} trap which is a hybrid between
free-space and surface geometries.  This trap is both deep and
tightly confining while consuming relatively little power.  The
trap has allowed for successful evaporation of a sample $^7$Li in
the $|F=2,m_F=2 \rangle$ ground state to quantum degeneracy.  As
evaporative cooling of $^7$Li is challenging due to its relatively
small scattering length~\cite{LiBEC_hulet,French_lithium}, we
expect this trap to scale well to evaporation of heavier alkalies
such as Na and Rb.

The paper is organized as follows. We start by reviewing top level
design criteria for magnetic traps.  We then describe our
implementation. Finally we present data on efficient evaporation
to BEC.

\section{Magnetic trap design criteria}
The performance of a magnetic trap depends on both local and
global parameters.  The local parameters are the parameters near
the center of the trap, such as the gradient and curvature of the
potential energy.  The global parameters are the parameters of the
boundary of the trap, such as the trap depth and the trap volume.
Both global and local parameters are important for achieving
efficient evaporation and for obtaining large numbers of
degenerate atoms at the end of evaporative cooling.  These
parameters are affected by the size and geometry of the trap.

Let us first look at the local parameters. In a typical
Ioffe-Pritchard trap, the radial magnetic field gradient and the
axial magnetic field curvature scale as $I/r^2$ and $I/r^3$
respectively, where $I$ is the current flowing in the magnet wires
and $r$ is the distance from the wires to the center of the trap.
If we assume the current density in the wires is $j$ and that trap
dimensions scale linearly with $r$, then the current of the trap
will scale as $j r^2$, so the radial magnetic field gradient
scales as $j$ and the axial magnetic field curvature scales as $j
/r$.  In theory, a small trap only gains in the axial magnetic
field curvature.  In reality, it is hard for a large trap to
maintain the same current density as a small trap since the power
dissipation scales as $j^2 r^3$ while the temperature drop with
respect to the heat sink increases as $r^2$. As a consequence,
increasing the size of the trap requires either a compromise in
the current density or the overhead of water cooling.  In
addition, as solid conductors would imply currents so large as to
be unpractical for power supplies and current leads, large
magnetic traps are realized with multiple windings, with wire
isolation further reducing the effective current density. In
micro-traps the current density in the wires can be orders of
magnitude larger than in traditional magnetic traps due the close
proximity of the conductor to the heat sink.  In the end, in order
to optimize local parameters, a small trap is preferred over a
large trap.

Now let us look at the global trap parameters.  The trap depth of
an Ioffe-Pritchard trap, which is determined by one of the 6
saddle points (4 in the radial direction and 2 in the axial
direction), scale as $j r$. The trapping volume scales as $r^3$.
It is important that the trap depth is higher than the average
initial kinetic energy of the atomic ensemble in order to ensure
that a substantial fraction of atoms is initially confined in the
trap. Furthermore, the trapping volume should be comparable to the
size of the initial ensemble, or again a significant fraction of
the atoms will fail to be initially confined.  Small traps rely on
auxiliary coils to adiabatically compress the atom distribution.
The effectiveness of this compression is limited by two factors.
First, compression increases the ensemble temperature, thus
raising the required trap depth of the small trap.  Second, for a
quadruple-type potential, the most widely used for this purpose,
shrinking the size of the sample by a factor $n$ is achieved at
the cost of an $n^3$ higher current and an $n^6$ larger dissipated
power. As a consequence, while small traps feature better local
parameters, large traps are superior in terms of global
parameters.

Trap parameters are also affected by the geometric arrangement of
the magnet wires.  For example, the ``Z" trap \cite{atomchip2} has
a planar structure and is well suited for atom chip designs, but
it is not as effective as a standard Ioffe-Pritchard trap in terms
of optimizing local and global parameters:  in the radial
direction, with the same trap depth and trap volume, the ``Z" trap
has a factor of 4 weaker radial field gradient; in the axial
direction, the trap depth is a factor of $2\pi$ lower with the
same trapping volume.

\section{The mini-trap}
Based on the above considerations, we designed a trap for
evaporative cooling of $^7$Li seeking the best compromise between
the local and global parameters. A similar approach has been
followed by other groups recently for Rb~\cite{Schmiedmayer,Dan}.
However, Li puts tighter constraints on the design for three
reasons. First, Li does not support sub-Doppler laser cooling. As
a consequence Li atoms are almost an order of magnitude hotter
than Rb before loading the magnetic trap.  Second, the s-wave
elastic collision cross-section of Li is an order of magnitude
smaller~\cite{LiScatter} than that of Rb~\cite{RbScatter}. This,
combined with a high two-body loss rate, impedes the evaporative
cooling process.  Finally, since the elastic collision
cross-section drops with increased temperature~\cite{daliba}, the
use of adiabatic compression to increase the elastic collision
rate is ineffective.  When designing our mini-trap, typical Li MOT
parameters were used to set trap global parameters.  We then
attempted to maintain a tight electrode structure to maximize the
local parameters.  The resulting trap design is discussed below.

\subsection{Mini-trap construction}
\begin{figure}
  \includegraphics[scale=.3]{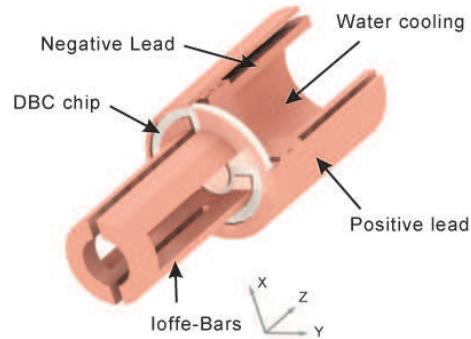}
  \caption{Assembled trap structure, including mini-trap coils, DBC interface chip,
  and current leads.}
  \label{assemble}
\end{figure}

A schematic view of the mini-trap assembly is shown in
Fig.~\ref{assemble}.  The trap is comprised of four parts: the
free-standing electrode structure, the DBC chip, the negative
current lead and the positive current lead.

\begin{figure}
  \begin{center}
  \includegraphics[scale=1.2]{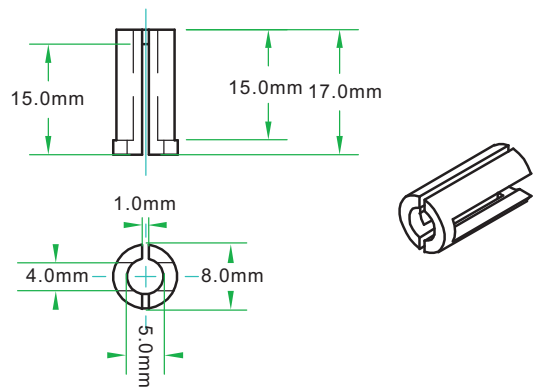}
  \end{center}
  \caption{Schematic illustration of the free-standing electrode structure.
  The coils are machined from OFHC copper and brazed to the DBC chip substrate
  shown in Figs. 1 and 3.}
  \label{copper}
\end{figure}

The first part, shown in Fig~\ref{copper}, is machined from a
solid piece of oxygen-free copper. We start from a 17 mm long
tube, with 5 mm inner diameter and 8 mm outer diameter.  Two
slits, respectively 4 mm and 1 mm wide, are cut orthogonally
through the tube in the longitudinal direction, stopping 2 mm from
opposite ends. In this manner, the partial rings at the ends of
the tube form pinch coils, while the remaining length is divided
into 4 Ioffe bars.  One end of the tube is brazed to a DBC ceramic
chip.

The second part is the DBC chip.  In the DBC process, Cu foil is
typically bonded to an alumina ceramic substrate in an N$_{2}$ gas
atmosphere at $\sim 1070^\circ$C. Reaction compounds such as
CuAlO$_{2}$ and CuAl$_{2}$O$_{4}$ form a strong bond in the
vicinity of interface between Cu and the alumina substrate. Traces
are then etched from the copper foil.  In our case, we designed
traces to route current through the Ioffe bars to the current
leads.  Compared to other mature micro-fabrication approaches such
as thin film processes and thick film processes, DBC provides
better thermal conductivity and allows for significantly larger
currents on the copper layer.

\begin{figure}
  \begin{center}
  \includegraphics[scale=0.3]{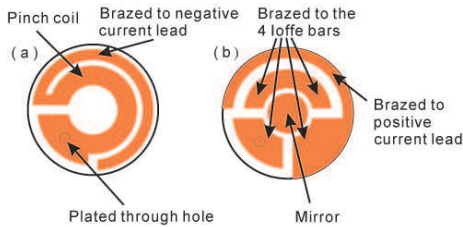}
  \end{center}
  \caption{(a) Back surface of the DBC chip. (b) Front surface of the DBC chip.
  The free-standing electrode structure illustrated
  in Fig. 1 is brazed on Cu pads on the
  side of the chip containing the central mirror.  A small via is used to
  run current from the negative current pad to the free-standing structure. }
  \label{chip}
\end{figure}

The schematics for the DBC chip are shown in Fig.~\ref{chip}.  The
small via on the pad provides an electrical conduit between both
sides of the chip.  On the front side of the chip, the outer trace
is welded to the positive lead of the power supply. In addition, a
Cu disc in the center is polished to form a mirror that can be
used to retro-reflect a laser beam (to achieve an optical lattice,
for example).   On the back side of the chip, the inner ring has
the same diameter as the copper piece and works as the other pinch
coil for the Ioffe-Pritchard trap. The outer ring is welded to the
negative lead of the power supply and the heat sink.

The third piece of the assembly is the negative current lead. The
middle of the lead is hollow and blind to run water cooling. The
fourth piece of the assembly is the positive current lead. It is
concentric with the negative current lead to avoid disturbing the
trap field.

The Ioffe bars and the current leads are welded to the DBC chip in
a high temperature ($700 ^\circ$C) vacuum furnace. The material
and brazing process are all ultra-high vacuum compatible.

\begin{figure*}
  \begin{center}
  \includegraphics[scale=0.8]{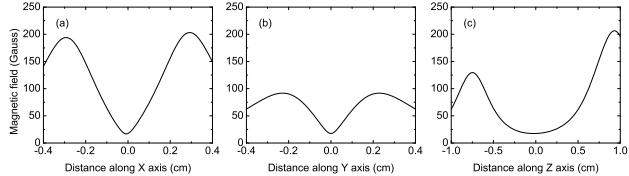}
  \end{center}
  \caption{Simulated fields for the mini-trap electrode structure for 100 A drive current.
  (a) Cross-section along x-axis. (b) Cross-section along y-axis.  (c) Cross-section along
  z-axis.}
  \label{IP_z}
\end{figure*}

\subsection{Predicted performance}
Calculations of the predicted magnetic field in the radial and
axial directions are shown in Fig.~\ref{IP_z}. All fields are
calculated for a 100 A trap current. The asymmetry in the radial
direction is due to the asymmetry of the Ioffe bars. Because we
seek a large numeric aperture in the x-direction for imaging, the
distance between the Ioffe bars in the x-direction is larger than
that in the y-direction. In the axial direction, the pinch coil at
the tip of the trap is only a partial circle, while on the back of
the chip, there is a full circle of copper trace.  As a result,
the field is weaker at the tip of the trap than close to the chip.
Away from the center, the field gradient in the radial direction
is about 800 G/cm in the x-direction and 400 G/cm in the
y-direction. In the area near the center of the trap, $\partial
B_{z} /\partial z=0$ so that $|\partial B_{x} /
\partial x|=|\partial B_{y} /\partial y|$ = 510 G/cm. In
the axial direction, the oscillation frequency is predicted to be
67 Hz. The trap depth of 70 G is due to the saddle point in the
y-direction.

\subsection{Evaporative cooling apparatus}
\begin{figure}
  \begin{center}
  \includegraphics[scale=0.3]{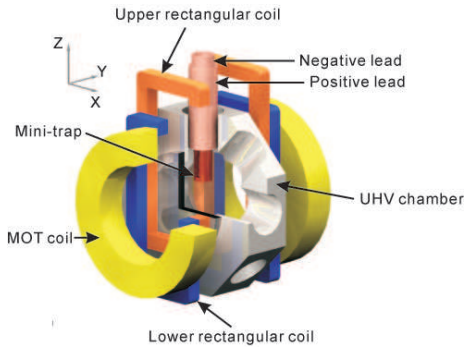}
  \end{center}
  \caption{Schematic illustration of the evaporative cooling apparatus.
  Shown are MOT coils, lower rectangular coils, upper rectangular coils,
  the negative lead of the mini-trap, the positive lead of the mini-trap,
  the mini-trap and the UHV vacuumm chamber. The trap is loaded from a
  a collimated Li atomic beam propagating along the x-axis.}
  \label{geometry}
\end{figure}

The mini-trap is integrated into an evaporative cooling apparatus
which is illustrated in Fig.~\ref{geometry}.  In brief, a
transversely cooled atomic beam loads a 3D MOT.  Atoms from the 3D
MOT are then optically pumped and transferred into the mini-trap,
located 2 cm above the MOT.  Auxiliary rectangular magnetic coils
are used to move the atoms from the MOT region to the mini-trap
region.  We detail each of these steps below.

The trap is loaded from a transversely laser cooled Li atomic
beam. Laser cooling is achieved by two pairs of zigzag broadband
laser beams \cite{Fabio}.  The 2D cooling region is separated from
the main vacuum chamber (which contains the mini-trap) by a
differential pumping tube.  Atoms entering the main chamber are
slowed and captured by a broadband 3D MOT \cite{brian}.
$~2\times10^9$ atoms are loaded into the MOT in 30 s.  The loading
rate is primarily determined by the Li oven temperature, which is
kept suitably low to maintain the high vacuum conditions required
for evaporative cooling.

After loading atoms into the 3D MOT, atoms are optically pumped by
a combination of hyperfine pumping and Zeeman pumping to the $|F =
2,m_f = 2 \rangle$ state.  Zeeman pumping is accomplished with a
circularly polarized laser which is tuned to $D_{1}$ F=2
$\rightarrow$ F'=2 transition. Unlike the $D_{2}$ line, the
$D_{1}$ line has a well resolved fine structure and the $|F = 2,
m_F = 2 \rangle$ ground state is a dark state under excitation
from circularly polarized pumping light.  A bias field is pulsed
on during the Zeeman pumping interval.

Following the optical pumping sequence, $8\times10^8$ atoms are
held in a magnetic quadrupole trap formed from the same coils used
for the MOT.  Atoms are then transferred from the  quadruple trap
by ramping down the current in the MOT coils and ramping up the
current in the lower rectangular coils in 50 ms.  In this step,
the center of the atom cloud does not change but the shape of the
cloud becomes elongated.  Next we ramp down the current in the
lower rectangular coils and ramp up the current in the upper
rectangular coils in 70 ms.  After this step, the elongated atom
cloud is in the center of the mini-trap, 2 cm above the center of
the MOT.

It is important to transfer the atoms from the quadruple trap to
the mini-trap with a minimum loss of phase space density.  In
order to minimize heating during the final transfer step, we
implemented a semi-adiabatic transfer scheme.  The idea is as
follows.  The radial confinement of the mini-trap is more than 2
orders of magnitude stronger than the axial confinement. We switch
off the rectangular transfer coils and switch on the mini-trap in
a time scale that is much slower than the radial oscillation
period but much faster than the axial oscillation period. In this
case, we achieve adiabatic transfer in the radial direction, and
therefore only need to match the trap geometry in the axial
direction (which can be accomplished by appropriate design of the
rectangular transfer coils).  We implemented the semi-adiabatic
transfer by ramping down the current in the upper rectangular
coils and ramping up the current in the mini-trap in 2 ms.  Using
this method, we achieved a transfer efficiency of 25\% from the
rectangular coils into the mini-trap. 100 msec after the transfer,
roughly $2\times10^8$ atoms remain in the mini-trap.

Due to the minimum loss of phase space density during the transfer
process, we achieved efficient evaporative cooling without an
additional Doppler cooling stage in the Ioffe-Pritchard trap
\cite{French_lithium}. To test the effectiveness of the
semi-adiabatic transfer, we reversed the current flow in the
rectangular transfer coils.  Although the atoms experience the
same potential before and after the transfer, the transfer is no
longer semi-adiabatic, and nearly all atoms are lost from the
mini-trap in just 2 s.

\section{Trap characterization}
We characterized the trap through RF and Zeeman spectroscopy and
parametric heating measurements, as described below.

The magnetic field in the center of the trap was measured using RF
spectroscopy.  For these measurements -- and also for evaporative
cooling -- we employed a swept RF source to eject atoms from the
trap at energies determined by the RF frequency.  Atoms were
ejected using transitions to the untrapped $|F=1, m_{F}=1 \rangle$
state.  When the applied RF frequency $\nu_{cut}$ reaches

\begin{equation}\label{bottom}
h(\nu_{cut}-\nu_{hfs})=(m_F g_F -m_F'g_F')\mu _{B}B_0,
\end{equation}

\noindent all atoms are ejected from the trap.  Here $\nu_{hfs}$ =
803.5 MHz is the groundstate hyperfine interval, $B_0$ is the
magnetic field at the trap bottom, and $g_F$ is the Land\'e
g-factor (primes distinguish between the initial and final state).

\begin{figure}
  \begin{center}
  \includegraphics[scale=0.7]{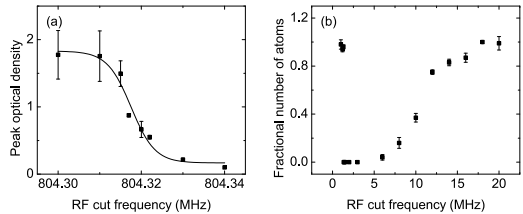}
  \end{center}
  \caption{(a) RF spectroscopy measurement of
  the trap bottom $B_0$ for a trap current of 100 A.  (b) Zeeman spectroscopy determination of the
  of trap depth for a trap current of 36 A.}
  \label{rfcut}
\end{figure}

For our measurements, we first prepared a 6 $\mu$K sample by
sweeping the RF frequency from 980 MHz down to 805 MHz. We then
applied a constant RF frequency cut for 10 s. Fig. \ref{rfcut}(a)
shows the number of atoms remaining in the trap as a function of
this frequency.  The sudden jump in atom number was used to
measure the magnetic field at the bottom of the trap. Our trap
bottom measurement has a frequency resolution of better than 10
kHz.

At 100 A, the bottom of the trap was measured to be 7 G,
significantly lower than the calculated value of 17 G.  We
confirmed this through direct measurement of the magnetic field of
a to-scale model of the trap.  We believe this discrepancy can be
explained by consideration of non-uniformities in the current
distribution in the leads on the DBC chip.   Finally, we verified
that the magnetic field at the trap bottom scaled linearly with
the trap current.

Because the minimum field resulting from the mini-trap alone is
relatively low, it can be precisely manipulated with an external
bias coil.  Tuning of the field minimum in this way can be used to
control the radial curvature of the trap.  For the evaporative
cooling demonstration described below, we have added an extra
field so that the overall trap minimum is 0.4 G.

We used Zeeman spectroscopy to measure the depth of the trap. For
these measurements, we confined atoms in a relatively weak 36 A
trap, in the $|F=1,m_F = -1 \rangle$ state, in order to ensure
full filling of the trap.  We then applied a 4 sec, single
frequency, excitation to drive Zeeman transitions to the $|F=1,m_F
= 0 \rangle$ untrapped state.  We subsequently measured the number
of atoms remaining in the trap, as shown in Fig. \ref{rfcut}(b).
From this data we inferred a trap depth of 66 G when results were
scaled to 100 A operating currents. This value is in good
agreement with our simulated values.

\begin{figure}
  \begin{center}
  \includegraphics[scale=0.9]{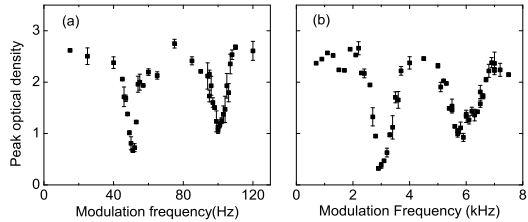}
  \end{center}
  \caption{ Optical depth vs. parametric drive frequency for
  determination of the radial and axial trap frequencies
  (a) Axial resonances, corresponding to $\omega_{\perp}=2\pi\times$
  3 kHz.
  (b) Radial resonances, corresponding to
  $\omega_{\parallel}=2\pi\times50$Hz.}
  \label{para}
\end{figure}

The local parameters of the mini-trap were measured by parametric
heating.  We parametrically heated atoms with an additional audio
frequency current (3 A) which was summed with the DC trap current.
To make the measurement we first prepared a 6 $\mu$K ensemble with
an RF sweep to 805 MHz.  After preparing this ensemble we turned
on the parametric excitation for a 10 sec interval. Following the
excitation interval we measured, using absorption imaging, the
optical depth at the trap center.  When the modulation frequency
$\omega$ and the atom oscillation frequencies $\omega_{\perp}$
(radial frequency), $\omega_{\parallel}$ (axial frequency) satisfy
the resonance conditions $\omega=2\omega_{\perp}/n$ or
$\omega=2\omega_{\parallel}/n$ (n integer), the parametric process
heats the atom ensemble and causes a reduction in the peak optical
density. The $n = 1$ and $n = 2$ resonances for a 100 A DC trap
current and center bias field of 0.4 G are shown in
Fig.~\ref{para}.

From Fig.~\ref{para}, we find a trap axial oscillation frequency of
$\omega_{\parallel}=2\pi\times$50 Hz and radial oscillation
frequency of $\omega_{\perp}=2\pi\times$3 kHz. From

\begin{equation}
\frac{\partial^{2} B}{\partial z^{2}} =
\frac{m_{Li}\omega_{\parallel}^2}{m_{F}g_{F}\mu_{B}},
\end{equation}

\noindent we calculate that the axial curvature is 120 G/cm$^2$
(m$_{Li}$ is the Li atomic mass). Similarly, in the radial
direction, we obtain $\partial ^{2} B/\partial r
^{2}=4.4\times10^5$ G/cm$^2$. From

\begin{equation}
\frac{\partial^{2} B}{\partial r^{2}}= \frac{(\partial B/\partial
r)^2}{B_{0}}-\frac{1}{2}\frac{\partial^{2} B}{\partial z^{2}},
\end{equation}

\noindent and $B_{0}=0.4$ G, we find the radial magnetic field
gradient $\partial B/\partial r = 420$ G/cm.  These results are in
reasonable agreement with the predicted fields.

\section{Evaporative cooling to BEC}
\begin{figure}
  \begin{center}
  \includegraphics[scale=0.5]{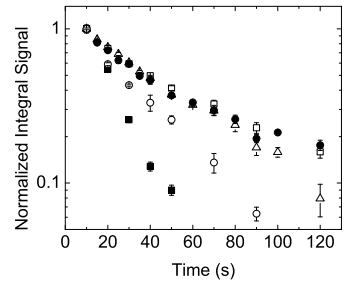}
  \end{center}
  \caption{Trap lifetime vs. current.  The fitted lifetimes are 87 s, 68 s, 54 s, 30 s
  and 15 s for currents of 78 A (filled circles), 100 A (open squares),
  120 A (triangles), 135 A (open circles),
  and 148 A (filled squares) respectively.}
  \label{comparecurrent}
\end{figure}

In order to optimize parameters for evaporative cooling, we
measured the trap lifetime as a function of trap current, as shown
in Fig.~\ref{comparecurrent}.  Our trap achieved a lifetime of 87
s at 78 A, limited by the vacuum in the main chamber. The lifetime
decreased with higher mini-trap currents, presumably due to
degradation of the vacuum due to resistive heating of the trap
electrodes.  However, at 120 A we still achieved a trap lifetime
of 60 sec, which was sufficient to support effective evaporative
cooling.  At this current, the current density was 35 A/mm$^2$ in
the Ioffe bars and as high as 200 A/mm$^2$ in the pinch coil on
the DBC chip.

\begin{figure}
  \begin{center}
  \includegraphics[scale=1.0]{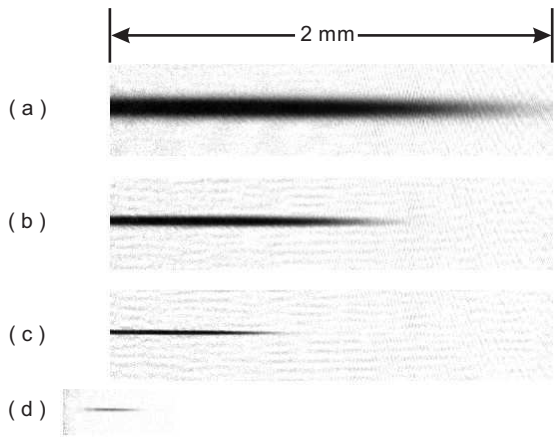}
  \end{center}
  \caption{Absorption images for final RF sweep values of
  (a) 816 MHz (1x10$^7$ atoms, 75 $\mu$K), (b) 808 MHz
  (3x10$^6$ atoms, 22 $\mu$K), (c) 805.4 MHz (8.4x10$^5$ atoms, 6.5 $\mu$K)
  and (d) 804.48 MHz (6x10$^4$ atoms, 1.1 $\mu$K).  Image (d)
  is at the BEC threshold.  The trap is only partially resolved in
  images (a)-(c) due to incomplete optical access.  For these images
  we crop the image at the center of the trap. }
  \label{absorp}
\end{figure}

To evaporatively cool trapped atoms we applied a 35 sec RF sweep
from an initial RF frequency of 980 MHz to a final RF frequency
close to $\nu_{hfs}$.  The sweep consisted of piecewise linear
steps which were independently optimized in order to maximize
phase space density at the end of the RF sweep.  We inferred phase
space density from absorptive imaging measurements of atom number
and temperature, and the measured trap frequencies. Our imaging
system had a diffraction limited resolution of 4 $\mu$m. Typical
absorption images of evaporatively cooled ensembles are shown in
Fig.~\ref{absorp}.

We extracted number and temperature from these images using the
following procedure.  First, the axial profile of the cloud is
fitted to a Gaussian distribution and the size
$\sigma_{\parallel}$ is used to deduce the temperature from

\begin{equation}
T=m_{Li}\omega_{\parallel}^2\sigma_{\parallel}^2/k_{B},
\end{equation}
where $\omega_{\parallel}$ is the measured axial oscillation
frequency. The peak density $n$ is deduced from the central
optical density together with the radial size of the cloud
$\sigma_{\perp}$. The central phase space density is
$D=n\lambda^3$, where $\lambda=\sqrt{2\pi \hbar^2/m_{Li}k_{B}T}$
is the thermal de Broglie wavelength.

\begin{figure}
  \begin{center}
  \includegraphics[scale=0.9]{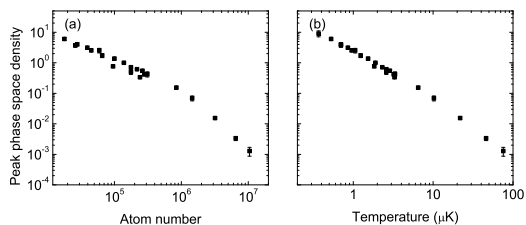}
  \end{center}
  \caption{(a) Peak phase space density vs. number of atoms. (b) Peak phase space
  density vs. temperature.}
  \label{psd}
\end{figure}

In Fig.~\ref{psd}(a), we plot the phase space density as a
function of atom number in the trap following the evaporative
cooling sweep.  The evaporation efficiency $\gamma \equiv
log(D_{f}/D_{i})/log(N_{i}/N_{f})$ (where $N_i$ and $N_f$ are the
initial and final numbers of atoms) is observed to be $\sim$ 2 for
phase space densities from $10^{-3}$ to $10^{-1}$, comparable to
previous Li BEC experiments~\cite{LiBEC_hulet,French_lithium}.
When the phase space density approaches the quantum degenerate
region, we observe a reduction in evaporation efficiency to $\sim$
1.  Here it is limited by two-body dipolar relaxation losses
\cite{Dipolar, Optimization}.  A plot of phase space density vs.
temperature is shown in Fig.~\ref{psd}(b).

Below the BEC transition, the negative scattering length limits
the number of condensed atoms to $<$ 600 atoms
\cite{BEClimitTheory,BEClimitExperiment}.  When the BEC fraction
exceeds this number, the BEC collapses and these atoms are ejected
from the trap.  In this work, we did not attempt to resolve the
BEC fraction near the phase transition.  However, we were able to
image very small ensembles of atoms ($\sim$ 300 atoms) in a nearly
pure condensate state.  We reach the BEC threshold with roughly
$6\times10^4$ atoms.

\section{Conclusion}
In conclusion, we have used a mm-scale mini-trap to achieve BEC in
$^7$Li.  We demonstrated a semi-adiabatic method to load the trap.
We characterized trap parameters using RF and Zeeman spectroscopy
and parametric heating measurements.  The trap dissipates only 7 W
power for operating parameters needed to achieve BEC.

It is interesting to consider extensions of this work to heavier
bosons such as Na or Rb, or for sympathetically cooled mixtures of
these species and other species ({\it e.g.} fermions). In these
cases, the efficacy of sub-Doppler laser cooling methods should
enable further miniaturization of the trap electrode structure. As
a result, power consumption could be an order of magnitude lower
($<$ 1 W).  In addition, evaporation times are expected to be much
faster than those obtained above, due to the significantly larger
elastic collision cross-sections of Rb or Na. The performance of
such a trap could compete favorably with other fast-evaporation
systems, such as micro-traps ~\cite{ChipBEC1} and all-optical
traps~\cite{all_optical}.  Ultimately, we expect this class of
traps could have broad impact for portable BEC systems.

\begin{acknowledgments}
This work was supported by DARPA and the NSF.  We thank Wayne
Rowlands, Fabio Peixoto, and Gilles Nogues for their assistance in
early stages of this work.
\end{acknowledgments}

\end{document}